\documentclass[preprint,superscriptaddress]{revtex4-1}
\usepackage{graphics,epsfig}
 
\usepackage{epstopdf}
\usepackage{graphicx}
 
\usepackage{dcolumn}
\usepackage{amsmath}
\newcommand{\be}{\begin{equation}}
\newcommand{\ee}{\end{equation}}
\newcommand{\bea}{\begin{eqnarray}}
\newcommand{\eea}{\end{eqnarray}}
\newcommand{\nn}{\nonumber}
\begin{document}

\title{How is the degree of freedom responsible for the entropy of the  black hole in BTZ spacetime?}
\author{Dharm Veer Singh}

\affiliation{Department of Physics, Institute of Applied Sciences and Humanities, GLA University Mathura, India}
\affiliation{Center for theoretical Physics, Jamia Millia Islamia, New Delhi}
\affiliation{Department of Physics, Banaras Hindu University, Varanasi}

\email{veerdsingh@gmail.com}
\author{Shobhit Sachan}
\affiliation{Department of Physics, Daulat Ram College, University of Delhi, New Delhi}
\affiliation{Department of Physics, Shri Ramswaroop Memorial University Barabanki, India.}
\email{shobhitsachan@gmail.com}

\begin{abstract}
 \noindent The entanglement entropy approach to study the dependence of  entropy upon the location of degrees of freedom  ($dof$) (near/far) from the horizon is discussed in this article. We try to understand the  physical deviation of  the area law for the excited  states by incorporating the logarithmic and power law corrections. We show that the  $dof$ near the horizon gives contribution to  the total entropy of the system in the ground state and away from the event horizon gives contribution to the excited state.
\end{abstract} 

\maketitle

Black holes are  non-singular solution of  Einstein field equation and  it behave as a thermodynamic object. It means we can easily study the black hole mechanics by apply laws of thermodynamics proposed by Bekenstein  \cite{JD,jd1,jd2,jd3} and Hawking \cite{SWH}. The temperature and entropy of the black holes are identified as the  surface gravity and area of  the horizon \cite{JD, SWH}. It radiates when quantum mechanical effect taken in account and this radiation is known as Hawking radiation. Many attempts have been made to  understand this radiation by studying the entropy of the black hole.  

Out of several methods used to explore the entropy in BTZ space time \cite{Carlip, Setare:2015pva,Setare:2011ec,Darabi:2010dp, Setare:2009hw, Cadoni:2008mw, Cadoni:2007ck,Setare:2006hq,Setare:2003hm, sud,sud1,sud2}, we focus our study to analyse the same using entanglement entropy approach which is most attractive candidate among these methods. The EE is a non-geometric method and in this approach we measures the  the quantum information due to the division of a system \cite{Bombelli,Srednicki,159,160,168}. The EE of other gravity theories are explored in Ref. \cite{l1,l2,l3}.

 The EE is one of the promising candidate to study the source of black hole entropy  and its sub-leading corrections (logarithmic and power law corrections) \cite{review,SOLO2,H2, DS2,DS5,Cadoni:2007vf,Cadoni:2009tk,Cadoni:2010vf}. It  arises due to the vacuum 
and thermal fluctuations \cite{152,s1,s2,s3,s4,s5,s6,s7}  in the vicinity of black hole space time. We  have made our attempts to study the scalar fields  propagating in the background of  BTZ  space time. We study the location of  $dof$ near and far from the event horizon of BTZ  space-time.  In this attempt to investigate the dependence of entropy on $dof$, two approaches have been  developed. The first approach associated with fundamental $dof$ related to string loops \cite {vafa, aa, adas} and other one associated with  quantum fields (scalar) propagating in the black hole space-time. 
We are using  the second  approach, one counts certain  $dof$ on the event horizon but can not say about the precise location relevant of the  $dof$.  The power law corrections arise due to thermal fluctuation \cite{152} in the presence of excited state (ES) \cite{161,DS3} and the logarithmic corrections arise due to the quantum fluctuations \cite {DS1, DS4}.

In this paper, we have made our attempts to study the  above-mentioned problems in more general framework, which may not be relevant to other approaches. We start our study by considering the entanglement between quantum fields lying inside and outside the black hole. The violation of area law \cite{161, DS3}  for ES can be understood by ascertaining the location of $dof$ far and near from the event horizon of the black hole, which leads to EE in these cases.
\newpage
\noindent The metric of the BTZ  space time is  given by the following line element \cite{MB}

\begin{equation}
{ds}^2=-\left(-M+\frac{r^2}{l^2}\right)\,dt^2+\left(-M+\frac{r^2}{l^2}\right)\,dr^2+r^2\,d\phi^2.
\end{equation}
The metric of the BTZ black hole can be written in terms of proper length
\be
{ds}^2=-k^2\,dt^2+\,d\rho^2+ l^2(k^2+M)d\phi^2,
\ee
where $r^2(\rho)=l^2(k^2+M)$ and  $M$ is the mass of the BTZ black hole. The scalar filed in background of BTZ black hole is 
 \be
S=-\frac{1}{2}\int dt\,\sqrt{-g}\,\,(g^{\mu\nu}\,(\partial_{\mu}\Phi\,\partial_{\nu}\Phi)-\mu^2\Phi^2),
\ee

using the separation of variables the field  $\Phi$   decomposed as,
\be
\Phi({t,\rho,\phi})=\sum_m\,\phi_m(t,\rho)\,e^{im\phi},
\ee
 and this decomposition of $\Phi$ manifest the cylindrical symmetry of the system. The scalar field in the presence of BTZ   space time  is
\be
S=-\frac{1}{2}\int dt\left[\frac{\sqrt{(k^2+M )}}{k}\,\dot{\Phi}_m^2+k\sqrt{k^2+M}(\partial_{\rho}\Phi_m^2)+\frac{k^2m^2}{k\sqrt{k^2+M}}\,\Phi_m^2\right],
\ee
 and the corresponding Hamiltonian is,
\be
H=\frac{1}{2}\int d\rho\,\tilde{\pi}_{m}^2(\rho)+\frac{1}{2}\int\,d\rho\,k\,\sqrt{k^2+M}\left(\partial_{\rho}\left(\frac{k}{\sqrt{k^2+M}}\right)\,\psi_m\right)^2+{\frac{{m^2k^2}}{{M+k^2}}}\psi_m^2,
\label{hamilt}
\ee
where
\begin{equation}
\psi_m(t,\rho)=\Big({\frac{{k^2}}{{k^2+M}}}\Big)^{1/4}\,\Phi_m(t,\rho),
\end{equation}
where $ {\tilde \pi}_m$ is canonical momentum corresponding to the field and it satisfy the following relation $[\phi_m(\rho), {\tilde \pi}_{m^{\prime}}(\rho^{\prime})]=\delta_{m,m^{\prime}}\delta(\rho-\rho^{\prime})$. The system can be discretized by the following replacements,
\begin{eqnarray}
 \rho\rightarrow(A-\frac{1}{2})a,~\qquad\qquad\delta(\rho-\rho ')\rightarrow \frac{\delta_{AB},
}{a}
\end{eqnarray}
where $A,B=1,2 ....N$ and ``$a$'' is ultra-violet cut-off .  The replacements of field is
\be
\psi_m(\rho)\rightarrow q^A,\qquad \tilde{\pi}_m(\rho)\rightarrow \frac{p_A}{a},\qquad V(\rho,\rho^{\prime})\rightarrow \frac{V_{AB}}{a^2}.
\ee
These replacements lead to the discretized Hamiltonian which is identical with the $N$ coupled Harmonic oscillator. It is written as,
\be
H=\sum_{A,B=1}^{N}\left[\frac{1}{2a}\delta^{AB}p_A\,p_B+\frac{1}{2}V_{AB}\,q_m^A\,q_m^B\right],
\ee

where $p_{A}=a\delta_{AB}\dot{q}^{B}$ is the canonical momentum corresponding to $q^{A}$. The interaction matrix elements $V_{AB}$ (where $A,B=1,2\ldots\ldots N$) is obtained by comparing the Eq. (3) and (8) which is written as \cite{DS1},
\begin{eqnarray}
V_{AB}=&&\frac{k_A}
{\sqrt{k^2_A+M}}\Big(k_{A+1/2}\sqrt{k^2_{A+1/2}+M}+k_{A-1/2}\sqrt{k^2_{A-1/2}+M}\Big)\delta_{A,B},\nn\\
&&\qquad-
2k_{A+1/2}\sqrt{k^2_{A+1/2}+M}\sqrt{\frac{k_A}{\sqrt{k^2_A+M}}}\sqrt{\frac{k_{A+1}}
{\sqrt{k^2_{A+1}+M}}}\delta_{A,B+1}\nn\\
&&\qquad\qquad\qquad\qquad\qquad+m^2\,\frac{k^2_A}{k^2_A+M}\psi_m^{A^2}\delta_{A,B}.
\label{int}
\end{eqnarray}
The diagonal and off diagonal terms of the matrix $V_{AB}$ can be identified from the equation (3). The various matrix elements of $V_{AB}$  matrix is written as,
\begin{eqnarray}
&&\Sigma_A^{(m)}=\frac{{k_{A}}}{\sqrt{(u^2_{A}+M)}}\Big(k_{A+\frac{1}{2}}\sqrt{k^2_{A+1/2}+M}- k_{A-1/2}\sqrt{[(k^2_{A-1/2}+M)}\Big)+m^2{\frac{{k^2_A}}{{(k^2_A+M )}}},\nn\\
&&\Delta_A~=~-k_{A+1/2}\sqrt{[(k^2_{A+1/2}+M)}\sqrt{\frac{{k_{A+1}}}{\sqrt{(k^2_{A+1}+M)}}}\sqrt{\frac{{k_{A}}}{\sqrt{(k^2_{A}+M)}}}.
\end{eqnarray}
These off-diagonal term of the matrix leads to interaction between the states in the same way as the nearest neighbors of the harmonic oscillator interacts. 

In order to understand how the  $dof$ near/far from the event horizon are contributing to the correction terms and how the correction arises/vanishes in both cases, we have to consider the state of quantum fields around the black hole. The  interaction matrix \ref{int} tells us about the location of degrees of freedom. The matrix form of the  $V_{AB}$ is,

\begin{center}
\begin{equation}
V_{(AB)}^m = \left( \begin{array}{lllllll} 
{\Sigma_1} & {} & {}& {} & {} & {} & {} \\
{} & {\Sigma_2} & {} & {} & {} & {} & {} \\
{} & {} & {\ddots} & {} & {} & {} & {} \\
{} & {} & {} & | \overline{{\Delta {A-2}}} & \overline{{\Sigma_{A-2}}} & \overline{{}}~~~~~~~| & {} \\
{} & {} & {} & | {{\Delta_{A-1}}} & {{\Sigma_{A-1}}} & {{\Delta_A}}~~~| & {} \\
{} & {} & {} & |\underline{\null} {} & \underline{{\Sigma_A}} & \underline{{\Delta_{A+1}}} | & {} \\
{} & {} & {} & {} & {} & {} & {\ddots} \\
\end{array} \right) \, .\label{int1}
\end{equation}
\end{center}
where the matrix element in the bracket is called the window. 
The percentage contribution of the entropy as a function of $q$ by using the following relation,
\begin{equation}
pc (q) ~=~ \frac{S (q, { fixed~} d)}{S_{ total}} \times 100 \, ,
\label{eqn11}
\end{equation}
where $S_{\rm total}$ is the total entropy of the system where  $i,j= 0...........N$. The entropy of the GS  are given by \cite{DS3}
\be
S=\sum_{i=1}^{N-n_{B}}S_{i}~~~ \mathrm{with}~~~ S_{i}=-\frac{\nu_i}{1-\nu_i}ln\,\nu_i-ln\,(1-\nu_{i}),
\ee
where 
\be
\nu_i=\frac{1}{\lambda_{i}}(\sqrt{1+\lambda_{i}}-1)^2,\qquad\qquad 0<\nu_i<1.
\ee
The Eq. (\ref{eqn11}) shows percentage contribution
of the total entropy which is the function  of window position ($q$).   In our calculations we have  taken fix value of $N$ and $n$, i.e $N=300$ and $n=100,~150$. Studying the Eq.  (\ref{eqn11}), we can say that,
\begin{itemize}
\item The percentage contribution of interaction term to the entropy is zero (from von Neumann entropy relation). The presence of interaction term raises the entropy significantly. We also observe that inclusion of  $ \it dof$ inside and outside the event horizon contributes to entropy.
\item The percentage contribution of entropy depends upon the window position. In the absence of any interaction,  peak is placed symmetrically inside and outside the horizon. This observation suggests that the maximum contribution to the entropy comes from those $ \it dof$ which are near to the event horizon.
\item For the excited state, the peak lowers as we increase the excitation number $o$. This suggests that, with increasing the excitation number ($o$), the significant contribution comes from the  $ \it dof$ far away from the event  horizon.
\end{itemize}

From the above discussion, we confirm that the entanglement between  $ \it dof$  inside and outside the  event horizon, contributes to the entropy. The  contribution to the entropy is more from the  $ \it dof$ near  the event horizon and they decrease with the increasing  excitation number.

\begin{figure}[ht]
\begin{tabular}{c c }
\includegraphics[width=8cm,height=6.5cm]{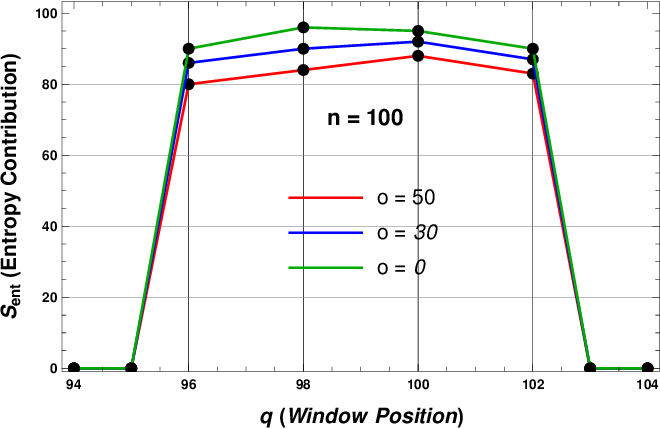}
\includegraphics[width=8cm,height=6.5cm]{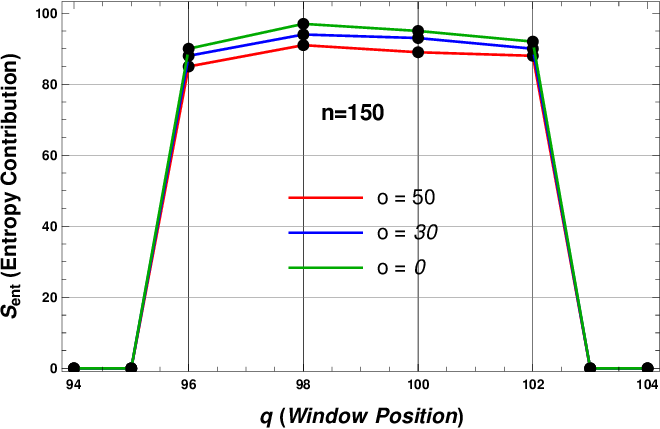}
\end{tabular}
\caption{Plot of total entropy as a function of window position for fixed window size $N=300$, $n = 100$ and $150$, for the ground state and excited state with excitation number $o=30$ and $50$.}
\label{fig:en1}
\end{figure}
 Now, we study how  $dof$ contribute to the entropy as a function of window width.  The percentage contribution to the EE for every window width $d$ can be calculate from the relation, 
\begin{equation}
pc(d)=\frac{S(d)}{S_{total}}\times 100.
\end{equation} 

\begin{figure} [h]
\begin{tabular}{c c}
\includegraphics[width=8cm,height=6.5cm]{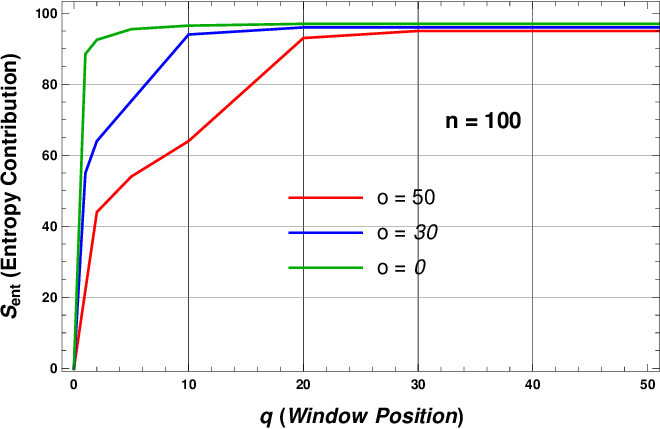}
\includegraphics[width=8cm,height=6.5cm]{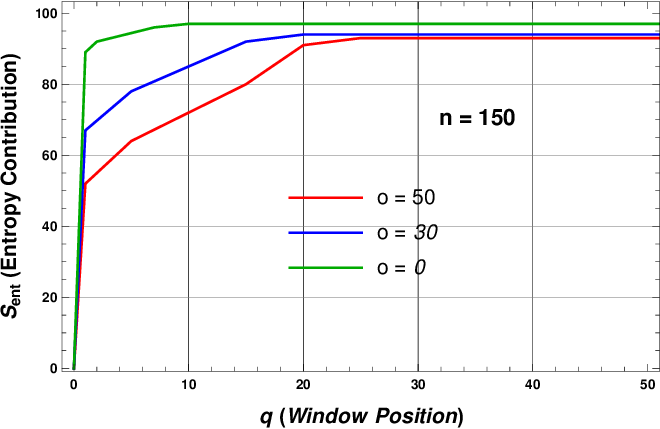}
\end{tabular}
\caption{Plot of the percentage contribution of $S_{ent}(t)$ for the
GS and ES, $N=300$, $n = 100$ and $150$ with excitation number  $o = 30$ and
$50$.}
\label{fig:en2}
\end{figure}

The effect of  $ \it dof$ on the entropy of the ES as shown in  Fig. 1.
We plot the graph of the percentage contribution of entropy with the window width in $d$ and the Fig. 2 for the GS, ES with $n=100$ and 150 for fixed $ N$.  This graph shows that for the GS follow the area law at the small value of the window width $d=5$, but for the ES it is recovered at the higher value of $d=20$ for the excitation number $o=30$ and $d=25$ for the excitation number $o=50$. From this, we can  conclude that the contribution of the  $dof$ near the event  horizon is more for the ES and increases with excitation number ($o$). The entropy depends upon the location of  $ \it dof$  for  ES. The logarithmic correction does not significantly affect the ES but present in GS \cite{DS1, MS}. The logarithmic corrections are present only in case of small black holes (the ES also contribute to the entropy on the form of power law corrections), but for the large limit, area law holds (see in Fig. 1 and Fig. 2). The logarithmic corrections in the entropy arise due to the high energy quantum fluctuations of fields near the horizon. These quantum fluctuations are small for macroscopic black holes and the leading term, which describe the situation of  thermal fluctuations are averaged out.

 We studied the location of  $dof$ near/far from the horizon. We have shown that how the  $dof$ are responsible for the entropy. The behavior of the  $dof$ are depicted in Fig. 1 and Fig. 2. We can see clearly from the Fig. 1 that the peak is symmetric inside and outside the horizon for the ground state but for the ES the peak diminishes with increasing excitation number $o=30, 50$. The peak increase when the window is near from the event horizon and decrease when far from the event horizon. From this observation, we can say clearly the  $dof$ are responsible for entropy and contribute more for the GS rather than ES. The Fig. 2 shows that the GS entropy and excited state entropy plots have similar behavior at large window width (position of  $dof$) but the effect is clearly visible at small window width. This is related to power law correction which arises due to the thermal fluctuations applicable near the event horizon, where a large number of the field modes ( $dof$) are present. These results would be more significant in the case of higher dimensional black holes. One can also study the dependence of entropy on  $dof$ for fermions and gauge fields in BTZ  space-time and non-singular black holes (where the entropy does not follow the area law) \cite{Singh:2017qur,Singh:2017bwj,Singh:2019tgw,Singh:2022xgi,Singh:2022tlo,Singh:2020xju}.

\section*{Acknowledgments}
DVS  acknowledges the UGC, India, for financial support under  DSKPDF scheme (Grant No.: BSR/2015-16/PH/0014).


\begin{thebibliography}{99}

\bibitem{JD}
J.~D.~Bekenstein, `` Black Holes and Entropy''  Phy. Rev. D {\bf 7}, 2333 (1973).
\bibitem{jd1}
 J.~D.~Bekenstein, ``Black holes and the second law'' Lett. al Nuovo Ciemnto 4 (1972) 15.
 \bibitem{jd2}
  J.~D.~Bekenstein,  ``Generalized second law of thermodynamics in black-hole physics'' Phy. Rev. D {\bf 9}, 3292 (1974).
  \bibitem{jd3}
   J.~D.~Bekenstein,  ``Statistical black-hole thermodynamics''    Phy. Rev. D {\bf 12}, 3077 (1975).

\bibitem{SWH} S.~W. ~Hawking, ``Black holes and thermodynamics '' Phy. Rev. D {\bf 13}, 191 (1976).


\bibitem{Carlip} 
S. ~Carlip,
{``The (2+1)-Dimensional Black Hole"}
{Class. Quantum Gravity}
{\bf 12}(1995) 2853.

\bibitem{Setare:2015pva}
M.~R.~Setare and H.~Adami,
``Entropy formula of black holes in minimal massive gravity and its application for BTZ black holes,''
Phys. Rev. D \textbf{91} (2015) no.10, 104039.



\bibitem{Setare:2011ec}
M.~R.~Setare and V.~Kamali,
``Correspondence between the contracted BTZ solution of cosmological topological massive gravity and two-dimensional Galilean conformal algebra,''
Class. Quant. Grav. \textbf{28} (2011), 215004.

\bibitem{Darabi:2010dp}
F.~Darabi, M.~Jamil and M.~R.~Setare,
``Self-gravitational corrections to the Cardy-Verlinde formula of charged BTZ black hole,''
Mod. Phys. Lett. A \textbf{26} (2011), 1047-1057.

\bibitem{Setare:2009hw}
M.~R.~Setare and M.~Jamil,
``The Cardy-Verlinde Formula and Entropy of the Charged Rotating BTZ black Hole,''
Phys. Lett. B \textbf{681} (2009), 469-471.

\bibitem{Cadoni:2008mw}
M.~Cadoni and M.~R.~Setare,
``Near-horizon limit of the charged BTZ black hole and AdS(2) quantum gravity,''
JHEP \textbf{07} (2008), 131.

\bibitem{Cadoni:2007ck}
M.~Cadoni, M.~Melis and M.~R.~Setare,
``Microscopic entropy of the charged BTZ black hole,''
Class. Quant. Grav. \textbf{25} (2008), 195022.

\bibitem{Setare:2006hq}
M.~R.~Setare,
``Gauge and gravitational anomalies and Hawking radiation of rotating BTZ black holes,''
Eur. Phys. J. C \textbf{49} (2007), 865-868.

\bibitem{Setare:2003hm}
M.~R.~Setare,
``Nonrotating BTZ black hole area spectrum from quasinormal modes,''
Class. Quant. Grav. \textbf{21} (2004), 1453-1458.


\bibitem{sud}S. Upadhyay, N,-ul-islam, P. A. Ganai, "A modified thermodynamics of rotating and charged BTZ black hole", JHAP \textbf{2}, 25 (2022).
\bibitem{sud1}N.-ul Islam, P. A. Ganai, S. Upadhyay, ``Thermal fluctuations to thermodynamics of non-rotating BTZ black
hole", Prog. Theor. Exp. Phys. \textbf{103B06}
(2019).
\bibitem{sud2}S. H. Hendi, S. Panahiyan, S. Upadhyay, B. E. Panah, ``Charged BTZ black holes in the context of massive gravity’s rainbow", Phys. Rev. D \textbf{95},
084036 (2017).

\bibitem{Bombelli}
L. Bombelli, R. K. Koul, J. Lee and R. D. Sorkin,
``Quantum Source of entropy for Black Hole"
Phy. Rev. D {\bf 34}, 373 (1986).

\bibitem{Srednicki} 
M. Srednicki, 
{``Entropy and Area" }
 {Phy. Rev. Lett.}{\bf 71} 666 (1993).
\bibitem{159}
S.~Das and S.~Shankaranarayanan,``How robust is the entanglement entropy: Area relation?,''
Phys.\ Rev.\ D {\bf 73} (2006) 121701.
\bibitem{160}
S Das, S. Shankaranarayanan, Sourav Sur,
 ``Black hole entropy from entanglement: A review" 
arXiv:0806.0402 [gr-qc].
\bibitem{168}
S. Das and S Shankarnarayanan,
 `` Where are the black hole entropy degree of freedom?," 
Classical and Quantum Gravity {\bf 24} 5299-5306 (2007).
\bibitem{l1}
L. Susskind,  Entanglement and Chaos in De Sitter Space Holography: An SYK Example. Journal of
Holography Applications in Physics, 2021; 1(1).
\bibitem{l2}
B. Kay,  Entanglement entropy and algebraic holography. Journal of Holography Applications in Physics,
2021; 1(1).
\bibitem{l3}
F.dos Santos,  Entanglement entropy in Horndeski gravity. Journal of Holography Applications in Physics,
2022; 3(1).
\bibitem{review}
 H.~Casini and M.~Huerta,
  ``Entanglement entropy in free quantum field theory,''
  J.\ Phys.\ A {\bf 42}, 504007 (2009).
\bibitem{SOLO2}
S.~N.~Solodukhin, 
{``Entanglement entropy of black holes,''}
Living Rev. Rel. {\bf 14} (2011) 8.
\bibitem{H2}
M.~Huerta,
{``Numerical Determination of the Entanglement Entropy for Free Fields in the Cylinder,''}
{Phys.\ Lett.\ B} {\bf 710} (2012) 691.
\bibitem{DS2}
D V Singh and S Siwach 
{``Thermodynamics of BTZ Black Hole and Entanglement Entropy''}
{Journal of phys. Conf. Series} {\bf 481} (2014) 012014.
\bibitem{DS5}
D.~V.~Singh and S. Sachan,
  ``Logarithmic Corrections to the Entropy of Scalar Field in BTZ Black Hole Space-time,''
  Int. Journal of Mod. Phys. D Vol. 26, (2017) 1750038.

 \bibitem{Cadoni:2007vf}
  M.~Cadoni, 
  ``Entanglement entropy of two-dimensional Anti-de Sitter black holes,''
  Phys.\ Lett.\ B {\bf 653} (2007) 434.
\bibitem{Cadoni:2009tk}
  M.~Cadoni and M.~Melis, 
  ``Holographic entanglement entropy of the BTZ black hole,''
  Found.\ Phys.\  {\bf 40} (2010) 638.
  \bibitem{Cadoni:2010vf}
M.~Cadoni and M.~Melis, 
``Entanglement Entropy of  AdS Black Holes,'' 
Entropy 2010, 12, 2244-2267.
\bibitem{152}
  A.~Chatterjee and P.~Majumdar,
  ``Black hole entropy: Quantum versus thermal fluctuations,''
 arXiv: gr-qc/0303030.
 
 \bibitem{s1}B. Pourhassan, H. Aounallah, M.Faizal, S. Upadhyay, S. Soroushfar, Y.
O. Aitenov, S. S. Wani,``Quantum thermodynamics of an M2-M5 brane system", JHEP \textbf{05}, 030 (2022).

  \bibitem{s2} B. Pourhassan, S. Upadhyay, ``Perturbed thermodynamics of charged black hole solution in Rastall
theory", Eur. Phys. J. Plus \textbf{136}, 311 (2021).

   \bibitem{s3} S. Upadhyay, B. Pourhassan, ``Logarithmic corrected Van der Waals black holes in higher dimensional
AdS space", Prog. Theor. Exp. Phys. \textbf{013B03} (2019).

   
    \bibitem{s4}S. Upadhyay, ``Leading-order corrections to charged rotating AdS black holes thermodynamics", Gen. Rel. Grav. \textbf{50}, 128 (2018).

     \bibitem{s5} S. Upadhyay, ``Quantum corrections to thermodynamics of quasitopological black holes", Phys. Lett. B \textbf{775}, 130 (2017).

      \bibitem{s6}S. Upadhyay, B. Pourhassan, H. Farahani, ``P-V criticality of first-order entropy corrected AdS black holes in massive
gravity", Phys. Rev. D \textbf{95}, 106014
(2017).

       \bibitem{s7}B. Pourhassan, M. Faizal, S. Upadhyay, L. A. Asfar, ``Thermal Fluctuations in a Hyperscaling Violation Background", Eur. Phys. J. C
77, 555 (2017).

 \bibitem{vafa}
A. Strominger, C. Vafa, ``Microscopic Origin of the Bekenstein-Hawking Entropy"  Phys. Lett. B {\bf 379}, 99, (1996).

\bibitem{aa}
  A.~Ashtekar, J.~Baez, A.~Corichi and K.~Krasnov,
``Quantum geometry and black hole entropy,''
  Phys.\ Rev.\ Lett.\  {\bf 80}, 904 (1998).
.
\bibitem{adas}
  A.~Dasgupta,
  ``Semi-classical quantisation of space-times with apparent horizons,''
  Class.\ Quant.\ Grav.\  {\bf 23}, 635 (2006).
         

\bibitem{161}
S.~Das, S.~Shankaranarayanan and S.~Sur, 
``Power-law corrections to entanglement entropy of black holes,'' 
Phys.\ Rev.\ D {\bf 77} (2008) 064013.
\bibitem{DS3}
D V Singh 
{``Power law corrections to BTZ Black hole Entropy ''} 
Int. J. Mod. Phys. D 24:1550001.

\bibitem{DS1}
D V Singh and S Siwach
{``Scalar Field in BTZ Black Hole Space-time and Entanglement Entropy''}, 
{Class. Quantum Grav.} {\bf 30} (2013) 235034.
\bibitem{DS4}
D V Singh and S Siwach
{``Fermion Field in BTZ Black Hole Space-time and Entanglement Entropy''}
Adv. High Energy Phys. \textbf{2015} (2015), 528762.


\bibitem{MB}
 M Banados, C Teitelboim and J Zanelli,
{``The Black Hole in Three Dimensional Spacetime"}
{ Phy. Rev. Lett.} {\bf 69}(1992) 1849.

\bibitem{MS} 
R.B.Mann and S.N.Solodukhin,
``Quantum scalar field on three-dimensional (BTZ) black hole instanton: Heat kernel, effective action and thermodynamics,
Phys.\ Rev.\ D {\bf 55}, 3622 (1997).


\bibitem{Singh:2017qur}
D.~V.~Singh and N.~K.~Singh,
``Anti-Evaporation of Bardeen de-Sitter Black Holes,''
Annals Phys. \textbf{383} (2017), 600-609.

\bibitem{Singh:2017bwj}
D.~V.~Singh, M.~S.~Ali and S.~G.~Ghosh,
``Noncommutative geometry inspired rotating black string,''
Int. J. Mod. Phys. D \textbf{27} (2018) no.12, 1850108.

\bibitem{Singh:2019tgw}
D.~V.~Singh and S.~Siwach,
``On Thermodynamics and Statistical Entropy of Bardeen Black Hole,''
[arXiv:1909.11529 [hep-th]].

\bibitem{Singh:2022tlo}
D.~V.~Singh, S.~Upadhyay and M.~S.~Ali,
``Rotating Lee\textendash{}Wick black hole and thermodynamics,''
Int. J. Mod. Phys. A \textbf{37} (2022) no.09, 2250049

\bibitem{Singh:2022xgi}
D.~V.~Singh, S.~G.~Ghosh and S.~D.~Maharaj,
``Exact nonsingular black holes and thermodynamics,''
Nucl. Phys. B \textbf{981} (2022), 115854
\bibitem{Singh:2020xju}
D.~V.~Singh and S.~Siwach,
``Thermodynamics and P-v criticality of Bardeen-AdS Black Hole in 4$D$ Einstein-Gauss-Bonnet Gravity,''
Phys. Lett. B \textbf{808} (2020), 135658
\end{thebibliography}
\end{document}